# The impact of MRI image quality on statistical and predictive analysis of voxel based morphology

*Felix Hoffstaedter[1,2], Nicolás Nieto[1,2], Simon B. Eickhoff[1,2], Kaustubh R. Patil[1,2]*

[1]Institute of Neuroscience and Medicine, Brain & Behaviour (INM-7), Research Centre Jülich, Jülich, Germany
[2]Institute of Systems Neuroscience, Medical Faculty, Heinrich Heine University Düsseldorf, Düsseldorf, Germany

## Abstract

Image Quality of MRI brain scans is strongly influenced by within scanner head movements and the resulting image artifacts alter derived measures like brain volume and cortical thickness. Automated image quality assessment is key to controlling for confounding effects of poor image quality. In this study, we systematically test for the influence of image quality on univariate statistics and machine learning classification. We analyzed group effects of sex/gender on local brain volume and made predictions of sex/gender using logistic regression, while correcting for brain size. From three large publicly available datasets, two age and sex-balanced samples were derived to test the generalizability of the effect for pooled sample sizes of n=760 and n=1094. Results of the Bonferroni corrected t-tests over 3747 gray matter features showed a strong influence of low-quality data on the ability to find significant sex/gender differences for the smaller sample. Increasing sample size and more so image quality showed a stark increase in detecting significant effects in univariate group comparisons. For the classification of sex/gender using logistic regression, both increasing sample size and image quality had a marginal effect on the Area under the Receiver Operating Characteristic Curve for most datasets and subsamples. Our results suggest a more stringent quality control for univariate approaches than for multivariate classification with a leaning towards higher quality for classical group statistics and bigger sample sizes for machine learning applications in neuroimaging.

*Index Terms* — group comparison, classification, structural MRI, image quality, sex/gender

## 1. Introduction

Quality of structural Magnetic Resonance Imaging (MRI) data significantly impacts derivative measures of brain morphology like local volume and cortical thickness [1]. Within scanner motion is the most common source of image artifacts and was shown to reduce both gray matter volume as well as cortical thickness estimates [2]. The evaluation of the image quality is also critical for accurate diagnosis of neurological disorders in clinical practice, and to this end, automated quality assurance models are under development [3].

Quality control is frequently conducted by trained individuals who visually inspect the reconstructed MRI scans and spot not only severe but also more subtle image artifacts with medium interrater convergence [1,2]. For large datasets of several hundreds to thousands of images, this labor-intensive procedure becomes less reliable and prone to errors. Common image processing tools, like Freesurfer and CAT12, include estimates of overall image quality that showed high correspondence with expert ratings [2,4]. It is a common practice to exclude structural MRI scans with severe movement artifacts, like "ringing" or "ghosting", or that show inferior quality compared to the analyzed sample. To avoid the influence of outliers with poor quality, a distance of 2-3 standard deviations from the average quality measure applied may be a useful rule of thumb. However, there is no generally accepted threshold of image quality to flag bad images for exclusion from further analysis, thus visual inspection of raw MRI scans is still considered a necessary step for images of minor quality.

Popular image processing tools are differentially affected by poor image quality with different amounts of images failing to be processed [5]. Overall, volumetric processing approaches seem more resilient to variances in image contrast as opposed to surface modeling approaches [1]. There is evidence that the sample-independent, weighted average image quality rating (IQR), derived by the popular image processing tool CAT12, is associated with FreeSurfer estimates of cortical thickness, surface area, and subcortical volumes [1]. Yet, the subsequent impact of image quality on traditional statistical modeling or machine learning performance is still not well understood. In this study, we aim to analyze the influence of MRI image quality on classical uni-variate statistics as well as multivariate prediction analysis of voxel-based morphology (VBM). To this end, we systematically investigated the effect of image quality on statistical sex/gender differences in local gray matter volume. Additionally, we analyzed the impact of image quality on the prediction of sex/gender from gray matter volume in a machine learning framework.

## 2. DATA AND METHODS

Our primary objective was to test the impact of image quality of structural MRI in the context of computational anatomy. We utilized three large publicly available datasets, from healthy participants covering the adult lifespan from 18 to 80 years. Specifically, we used the Southwest University Adult Lifespan Dataset (SALD, N = 494) [6], the enhanced Nathan Kline Institute Rockland Sample (eNKI, N = 818) [7], and the Cambridge Center for Ageing and Neuroscience sample (CamCAN, N = 651) [8]. Data was pre-processed using CAT 12.8.1 [9] with default settings, which yields for each T1 weighted image (1) an IQR value, as a quality measure, and (2) voxel-wise modulated gray matter volumes in Montreal Neurological Institute (MNI) template space. Then a whole-brain mask was employed to select 238,955 voxels before smoothing with a 4 mm full width at half maximum (FWHM) Gaussian kernel and resampling using linear interpolation to 8 mm spatial resolution yielding a total of 3747 features used in our analysis. The IQR is an absolute quality metric that is comparable over samples and computed as the weighted average of a noise-to-contrast ratio, an inhomogeneity-to-contrast ratio, and the image resolution [9]. Images are rated between 1-6 with 1 being excellent and <4 representing questionable quality including artifacts. In the samples used here, the average image quality was between 2-2.5 which can be considered as good to satisfactory quality [9].

As age represents the greatest source of variance in VBM measures, each dataset was balanced not only in terms of the target (sex/gender) but also in terms of age. This was achieved by stratified, balanced subsampling, dividing the age range (18-80) in equally distributed age bins, and retaining the same number of participants for each sex per bin. To generate sub-samples with different image quality, the participants retained in each age bin were selected according to their IQR. For high quality, participants with the lowest IQR were excluded, and the opposite was done to obtain low-quality subsamples. To generate a random quality sub-sample, participants were randomly sampled 20 times. In our experiments, two different numbers of age bins were used, 3 and 10. Using 10 age bins, 200 participants were retained from the SALD dataset and 280 were retained from eNKI and CamCAN. Using this number of bins the overlap of data between low and high-quality samples was minimized, but at the cost of only retaining 39% of available data (n=760). For 3 age bins, 336 participants were retained for SALD, whereas 426 from eNKI and 322 were retained for CamCAN. In this way, a bigger but more overlapped sample was obtained, retaining 56% of the data (n=1094).

For the traditional univariate statistical analysis, the complete pool of data was used for each number of age bins. The total intracranial volume (TIV) was linearly regressed out of the features and a Student's t-tests for sex/gender differences were applied. The resulting p-values were Bonferroni corrected and the number of significant tests ($p < 0.05$) was recorded per quality subsample.

For the machine learning analysis, a logistic regression model was used to predict sex/gender after leakage-free confound regression of TIV from the features. One model was trained for each dataset and sampling strategy and an additional model was trained using the pooled data. To address generalizability, a 5 times repeated 5-fold cross-validation approach was used and the Area under the Receiver Operating Characteristic Curve (AUC) was obtained on the test folds.

## 3. RESULTS

Mean IQR was similar for all datasets, with eNKI showing the best mean of 2.07 for the high-quality subsample, while CamCAN presented the worst average of 2.57 for the low-quality subsample (Table 1).

| IQR in subsamples of low, high, and random image quality (10 age bins) | | | | |
|---|---|---|---|---|
| | SALD | eNKI | CamCAN | Pooled data |
| low Q | 2.53 ± 0.23 | 2.49 ± 0.33 | 2.57 ± 0.32 | 2.53 ± 0.30 |
| random Q | 2.37 ± 0.21 | 2.26 ± 0.27 | 2.34 ± 0.29 | 2.33 ± 0.30 |
| high Q | 2.22 ± 0.12 | 2.07 ± 0.14 | 2.13 ± 0.12 | 2.33 ± 0.31 |
| IQR in subsamples of low, high, and random image quality (3 age bins) | | | | |
| | SALD | eNKI | CamCAN | Pooled data |
| low Q | 2.53 ± 0.23 | 2.49 ± 0.33 | 2.57 ± 0.32 | 2.38 ± 0.26 |
| random Q | 2.36 ± 0.21 | 2.25 ± 0.27 | 2.28 ± 0.25 | 2.29 ± 0.24 |
| high Q | 2.28 ± 0.15 | 2.07 ± 0.14 | 2.13 ± 0.12 | 2.29 ± 0.24 |

**Table 1.** Image quality over subsamples within each site

In the univariate analysis, for the low-quality images in the pooled subsamples with 10 age bins (n=760),

only 14 of the 3747 t-tests revealed significant sex differences (0.4%), while in the high-quality sub-samples 219 of the features were significant (5.8%) (Table 2). For random quality samples, 108 or 2.9% of the tests showed sex differences. For the larger sample, obtained with 3 age bins (n=1094), slightly better quality was observed (Table 1). In this case, the t-tests for low-quality data yielded 7.9% or 299 of features as significant, while the random and high-quality quality subsamples showed 12% or 450 and 19.5% or 731 features with significant effects of sex/gender, respectively.

| Number of significant features after Bonferroni corrected, 10 age bins. N [Median p-value of significant features] for pooled data | |
| --- | --- |
| low Q | 14 [7.49 e-06] |
| random Q (mean of 20 repeats) | 108 [4.12 e-06] |
| high Q | 219 [3.00 e-06] |
| Number of significant features Bonferroni corrected, 3 age bins. Mean [Median p-value of significant features] for pooled data | |
| low Q | 299 [2.79 e-06] |
| random Q (mean of 20 repeats) | 450 [1.94 e-06] |
| high Q | 731 [1.35 e-06] |

**Table 2.** The number of significant features of the Pooled data after Bonferroni correction.

For the machine learning analysis, AUC was generally lower in the SALD dataset, compared to the results obtained for eNKI or CamCAN, in all analyses. The poorest classification performance in SALD was obtained for models using the low-quality subsample, obtaining an AUC of 0.64-0.68 for 10 and 3 age bins, respectively (Fig. 1). For the same dataset, the classification performance showed the biggest improvement when using the high-quality subsamples, mainly in the 10 age bins, where the AUC increased 0.07, obtaining an AUC of 0.71. Models using random-quality data showed performance between those that used the low and high. For the pooled data, a similar trend as observed in the SALD dataset was obtained, where the models obtained better performance using higher data quality. For eNKI and CamCAN, all quality subsamples showed similar prediction accuracy, 0.77 for eNKI in both age bins and 0.74 and 0.70 for CamCAN using 10 and 3 age bins, respectively.

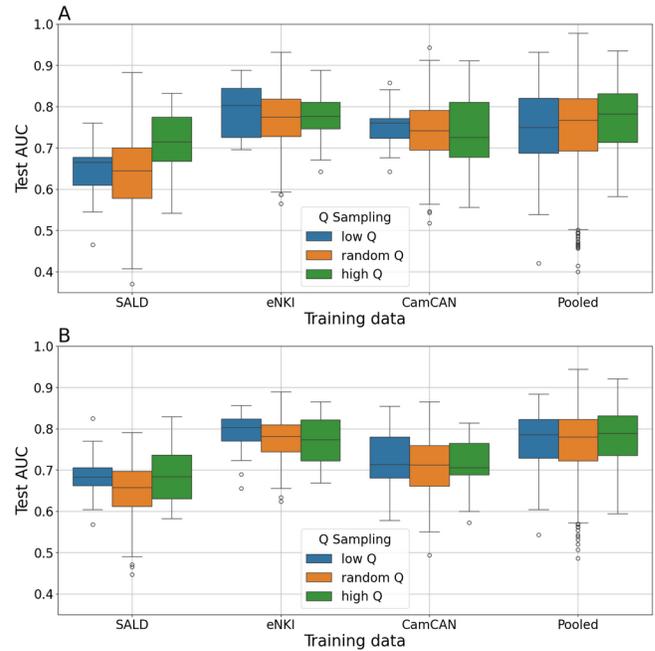

**Fig. 1.** AUC for sex/gender classification. A) AUC for subsamples with 10 age bins for low, random and high-quality images and overall for the pooled data. B) AUC for bigger subsamples with 3 age bins for each subsample and pooled data.

## 4. DISCUSSION

In this study, we explored the influence of MRI data quality on prototypical statistical analysis and machine-learning-based prediction. We used three different datasets to create subsamples of low, high, and random image quality for two subsampling strategies controlling the influence of age. Univariate effects of sex/gender were quantified with the number of significant t-tests over image features. To predict sex/gender, logistic regression models were trained within each and pooled datasets, and classification performance was compared between subsamples of image quality.

For univariate analyses, we found that using data with poorer quality resulted in much lower sensitivity to group differences for n=780. Increasing the sample size to n=1090 yielded a big increase in sex/gender effects emphasizing the influence of sample size on classical univariate analyses. With a limited amount of data, a focus on better image quality may increase the chance of detecting effects for classical statistical group comparisons.

Classification of sex/gender based on gray matter volume was less influenced by image quality or sample size in the current setup with acceptable image quality. It is timely to mention that the quality of the analyzed data were consistently good and only a few images presented a worryingly bad quality. Additionally, the similar results can be also explained by the overlapping between subjects retained in the different sampling strategies. Lastly, the inclusion of a broader range of image quality could represent the best option for machine learning approaches potentially yielding the biggest sample size without jeopardizing classification accuracy. Finally, the use of automatically generated image quality metrics, like the IQR, showed potential towards aiding, and ultimately replacing, visual inspection of structural MRI scans for large datasets.

## 5. ACKNOWLEDGMENTS

This work was partly supported by H2020 Research Infrastructures Grant EBRAIN-Health 101058516 and the Helmholtz Imaging Platform (project BrainShapes).

## 6. DATA AND CODE AVAILABILITY

Data used in the preparation of this manuscript are available publicly (might require registration and approval). The code used to generate the results and figures is available at https://github.com/N-Nieto/QC.

## 7. ETHICS STATEMENT

Ethical approval and informed consent were obtained locally for each study covering both participation and subsequent data sharing. The ethics proposals for the use and retrospective analyses of the datasets were approved by the Ethics Committee of the Medical Faculty at the Heinrich-Heine-University Düsseldorf.